\documentstyle[12pt]{article}
\def\tdfullput #1 #2 #3 #4 
  {\begin{figure}\vspace*{#3 cm}\vspace{12.3cm}
   \tdpsinput #1 #4 #3 #2
   \vspace{-4.0cm}              
   \end{figure}
   \typeout{TDINPUT: FIGURE \thefigure. produced with file #1}}
\def\tdfullfigure #1 #2 #3 #4 #5 #6
  {\begin{figure}\vspace*{#3 cm}\vspace{12.3cm}
   \tdpsinput #1 #6 #5 #3
   \vspace{-4.0cm}\vspace{#4 cm}
   \caption[]{#2}\end{figure}
   \typeout{TDINPUT: FIGURE \thefigure. produced with file #1}}
\def\tdpsinput#1 #2 #3 #4 {
   \special{ps::
-1 1 scale
-90 rotate
-1700 -2342 translate
#2 #2 scale
#4 -118 mul #3 -118 mul translate
3.5 3.5 scale
      }
   \special{ps: plotfile #1 asis}
   \special{ps:: endexecute
      }}
\begin{document}

\vspace*{2.5cm}

\begin{center}
{\Large {\sc Phase Transitions in Finite Nuclei and the}}

\vspace{2mm}

{\Large{\sc  Integer Nucleon Number Problem}}

\vspace{1.8cm}

R.F.Casten$^a$, Dimitri
Kusnezov$^{b,}$\footnote{dimitri@nst4.physics.yale.edu} 
and N.V.Zamfir$^{a,c,d}$\\

\vspace{1cm}

{\sl $^a$ WNSL, Yale University, New Haven, Connecticut 06520, USA \\
$^b$ Center for Theoretical Physics, Yale University, 
     New Haven, CT 06520-8120\\
$^c$  Clark University, Worcester, Massachusetts 01610, USA \\
$^d$   National Institute for Physics and Nuclear Engineering,\\
Bucharest-Magurele, Romania}

\vskip 1.2 cm

{\it To Appear in Physical Review Letters, June 1999}

\vspace{1.2cm}

\parbox{13.0cm}
{\begin{center}\large\sc ABSTRACT \end{center}
{\hspace*{0.3cm}
The study of spherical-deformed ground--state phase transitions
in finite nuclei as a function of N and Z is hindered by the
discrete values of the nucleon number.  A resolution of the
integer nucleon number problem, and evidence relating to phase
transitions in finite nuclei, are discussed from the
experimental point of view and interpreted within the framework
of the interacting boson model.
}}
\end{center}

\newpage

\setcounter{page}{2} 

It has been known for decades that nuclear properties change, often
dramatically, as a function of N and Z.  However, the possibility of true
nuclear {\it phase} coexistence and {\it phase} transitions in the evolution
of
nuclear structure as a function of nucleon number, in the sense of
conventional
condensed matter systems, has generally been discounted.  The reason has to do
with the finite nature of atomic nuclei and the fact that they contain integer
numbers of nucleons.   In order to discuss the concept of phase
transitions \cite{aa} one needs to identify a {\it control} parameter as
well
as an {\it order} parameter.  If the fluctuations in the order parameter are
small, that is, if the data all lie along a compact trajectory as a function
of
the control parameter, the existence of a phase transition would be signaled
by
a critical point where the order parameter is discontinuous.

The difficulty in applying this approach to nuclei lies in the
obvious fact that nuclei contain {\it integer} numbers of
nucleons and, therefore, their properties change {\it
discretely} as a function of N and Z.  This is illustrated in
fig.  1a where a typical collective observable in even-even
nuclei, the energy of the first 4$^+$ state, is plotted against
neutron number for the Sm (Z=62) isotopes.  Clearly, an
interesting (and well known) change in structure is occuring but
the abscissa values are, by definition, discrete.  Therefore,
regardless of what the data do, the integer nucleon number
requires that, at best, one can only connect adjacent points by
straight line segments and one can only define {\it differences}
in properties, not derivatives.

It is the purpose of this Letter to discuss phase
transitions in finite nuclei and to suggest a resolution of the
integer nucleon number problem.  When one
considers low energy nuclear structure, nuclear models offer the
flexibility of having one or more continuous parameters
that can serve as control parameters, allowing the study of
critical phenomena. The situation in actual nuclei is different.
In contrast to
phase transitions in a specific nucleus, where excitation energy (or
temperature) can be introduced as a control parameter, structure at low energy
evolves from nucleus to nucleus as a function of $N$
and $Z$, which we have seen are not useful as control parameters.
However, we will suggest an empirical quantity which is
continuous and which does correlate extremely well the
structural changes with N and Z across large regions of nuclei, and then apply
this approach to
the A = 150 region\cite{cc,dd,4a}.  Finally, a
theoretical analysis with the Interacting Boson Model (IBM)\cite{iba}
shows why sharp transition regions are indeed an expected feature of
structural evolution.

We noted in fig. 1 that the
data from a single element do not allow a discussion of phase transitional
behavior.  The data for collective nuclei in an entire region (50 $<$ Z $\leq$
66),
shown in fig. 1b, only exacerbate the problem.  The
abscissa values remain discrete, but now another prerequisite
for a phase transition disappears, namely, the absence of
fluctuations.  The scattering of the data obliterates any evidence of sharply
discontinuous behavior.

How can we get around this situation in finite nuclei?  A
possible answer, which we propose here, is to choose a
qualitatively different quantity to play the role of a control parameter, one
that is
at least potentially continuous.  Of course, even if the abscissa points
become
continuously distributed, we need to produce an approach in which the
fluctuations in the
data are small.  To do so, consider fig. 2a-c,
 which should be read as successive frames. In
this figure, the same E(4$^+_1$) data as in fig. 1 are plotted, not against N,
Z, or A, but against another collective observable, the energy, E(2$^+_1$), of
the first 2$^+$ state.  The top left panel is no better for discussing phase
transitional behavior than is fig. 1a.  The data points are discrete both
vertically and horizontally.  Although a nucleus can, in principle, have any
2$^+_1$ energy, the nuclei (Sm isotopes) in fig. 2a have specific and discrete
values since their properties change rapidly from isotope to isotope.


However, when the data for additional elements are added (fig. 2b,c)
we see a behavior that is qualitatively
different from fig. 1.  The distribution of points as a function of E(2$^+_1$)
successively fills in, yielding in fig. 2c a nearly continuous distribution.
Next, we note that
E(2$^+_1$) clearly correlates nuclear equilibrium
properties
extraordinarily well.  The data for
different elements lie along essentially {\it identical}
trajectories and thus the ensemble of data also lies along a
{\it single compact} curve, with very small fluctuations.  This thereby
enables
a discussion of the trajectory and a potential interpretation in terms of
phase transitions.
Of course, other observables that reflect the equilibrium configuration could
also have been
chosen.  E(2$^+_1$) is preferred, however, since it is well and accurately
known in a large
number of nuclei and is easy to measure in newly accessible nuclei.

Observables such as
E(2$^+_1$), E(4$^+_1$), separation energies  and other measures of structure
cannot, of course,
rigorously speaking, be considered as control parameters since they are not
independently
variable as is, for example, the temperature in a condensed matter system.
Nevertheless,
E(2$^+_1$), de facto, plays a similar role to a control parameter, and fig. 2c
suggests evidence
for true phase transitional behavior as seen by the nearly discontinuous
change in slope from
2.00 to 3.33 at a specific value of E(2$^+_1$) which we denote E$_c$(2$^+_1$).
  The expanded view of the rotational region in fig. 2d clearly shows the
different slope for these nuclei.  The change in slope occurs at
E$_c$(2$^+_1$) $\sim$ 120 keV, although this value may differ
in regions where other shells are filling.

Figure 2c itself is not a new result.  We have discussed the
E(4$^+_1$) - E(2$^+_1$) and related correlations in ref.\cite{czb} and
even broached the subject of phase transitional behavior.  The
correlation in fig.  2c has been discussed theoretically\cite{jolos} in the
context of the 1/N expansion method for the IBM,
and extended\cite{cho,buca} empirically to observables for intrinsic
excitations and to
odd A and odd-odd nuclei.  What is new here is the explicit
discussion of the {\it process} of reaching fig. 2c as a way of
resolving the finite nucleon number problem and identifying
phase transitional behavior.  Of course, since nuclei are finite systems, the
phase transition is smoothed out over a narrow region of 2$^+_1$ energies.

Our next point is to relate this phase transitional behavior to
the recent evidence\cite{dd,4a} for phase {\it coexistence} in
$^{152}$Sm (not merely the familiar shape coexistence originating from an
intruder state mechanism).  Theoretical analysis\cite{dd} of $^{152}$Sm points
to two co-existing phases, a deformed ground state band and a spherical
anharmonic vibrator spectrum built on the 0$^+_2$ level.  Phase
coexistence must occur at the critical point and so it is reassuring that
the 2$^+_1$ energy of $^{152}$Sm (121 keV) occurs at the same energy as
that where the change in slope from 2.00 to 3.33 occurs in
fig. 2c.

Thus, two different perspectives - phase coexistence in
$^{152}$Sm and the relation of yrast energies across the {\it region} of
nuclei - give evidence for a phase transition from spherical to deformed
structures, near A =
150.  Note that, while phase {\it coexistence} occurs in a specific nucleus
($^{152}$Sm), the
phase {\it transition} does not characterize a single
nucleus, or the isotopes of an element, but is a property of, and only
definable in terms of, an entire region.  We stress here that this type of phase
transition and phase coexistence is different from the shape coexistence picture
known in other regions \cite{wood,lhe}.  Here the structural changes develop
within the context of a single shell and do not involve an intruder state
mechanism.

Other observables reflect these rapid structural changes.  Two nucleon
separation energies,
S$_{2n}$, are sensitive to changes in equilibrium configurations.  In fig. 3a
we show the
empirical values of
S$_{2n}$  for the 50 $<$ Z $\leq$ 66 nuclei (for all collective
nuclei - those with R$_{4/2} >$ 2.05).  The S$_{2n}$ values
have a well known, essentially parallel, shift in values for each successive
Z. To compare values for different elements on the
same scale, we therefore shift the separation energies for each Z by a
constant amount chosen to give equal S$_{2n}$ values at N = 88 for the
N $>$ 82
shell and at N = 76 for the N $<$ 82 shell.

For N$>$82, the results for S$_{2n}$ in fig. 3a are as striking as for the
4$^+_1$ energies in fig. 2c.
The behavior is compact, with small fluctuations and a sharp
break in trajectory.  This break occurs at a slightly lower 2$^+_1$ energy
than the slope change for E(4$^+_1$) in fig. 2c.  Apparently,
the S$_{2n}$ values display a different dependence on the overall shape of the
nuclear potential than low spin yrast levels (which presumably are most
sensitive to the details of the potential near its minimum).  For the lighter
mass shell, N$<$82, there are greater fluctuations and a gradual structural
change but no evidence for a sharp phase transition.

In fig. 3b we show B(E2: 2$^+_1\rightarrow 0^+_1$) values as a
function of $S_{2n}$.  The sudden break of trend at a specific value of
S$_{2n}$ illustrates the
point that E(4$^+_1$) is not a unique measure of the structural transition:
$B(E2)$ values provide additional evidence for it.

The question arises as to {\it why}
nuclei should behave in this way.  As we have noted, it is easier to
look at phase transitional behavior in a model since the model parameters are
inherently continuous.  Here, we consider the IBM since the
analysis is simple but similar results characterize the
Geometric Collective Model (GCM) \cite{8a}.  We use
the
Extended CQF Hamiltonian\cite{wc,ltw}

\begin{equation}
H = \epsilon n_d - \kappa Q \cdot Q,
\end{equation}
where $Q = (s^{\dagger} \tilde{d} + d^{\dagger} s) +
\chi(d^{\dagger}
\tilde{d})^2 $, which has parameters $\epsilon$, $\kappa$, $\chi$
and the boson number $N_B$. Performing a systematic analysis of parameter
space,
IBM calculations of $E(4_1^+)$ versus $E(2_1^+)$ reproduce the results of fig.
2c:  they follow a
slope of 2.00 above E$_c$(2$^+_1$) for
virtually any choice of $\epsilon$, $\chi$, and N$_B$ that gives R$_{4/2}$ =
2.05 - 3.15 as long as
$\kappa$ is constant \cite{nvz}.  Indeed, the value of $\kappa$ determines
the intercept
$\epsilon_4$ [and E$_c$(2$^+_1$) as well].
Were the data different from fig. 2c (e.g., scattered, or
following a slope other than 2.00) the only way this behavior could be
reproduced would be to separately adjust
$\kappa$ for each nucleus, a clearly unlikely scenario, and one that is
inconsistent with microscopic analyses of IBM parameters \cite{16a}.

We now study the behavior of eq. (1) through
the intrinsic state formalism\cite{gk,dsi}, computing the energy surfaces
corresponding to different parameter values.  It is convenient, for
this purpose, to re-write eq. (1) with a control parameter
$\xi=(1+\epsilon/\kappa)^{-1}$. The resulting scaled Hamiltonian
has the form

\begin{equation}
H' = (1 - \xi)n_d - \xi Q \cdot Q.
\end{equation}

\noindent For $\chi$ = -$\sqrt{7}$/2, 0 $\leq$ $\xi$ $\leq$ 1 maps the
transition from U(5) to SU(3).

In fig. 4a we show the IBM energy surface corresponding to the classical limit of
eq. (2), for N$_B$ = 10
and $\chi$ = -$\sqrt{7}$/2, against $\xi$.
The figure immediately shows the key point, that the location of the minimum
in the energy,
$\beta_{min}$, changes suddenly, for these parameters, at a particular
$\xi$ value, from
$\beta_{min}$ = 0 to a large finite value, as indicated by
the dark line cutting
through the contour plot. There are
virtually no $\xi$ values for which intermediate $\beta_{min}$ values
result.  This is consistent with the specific calculations in
ref. \cite{dd}.  The IBM indicates that nuclei change abruptly
from near spherical to deformed at a critical value $\xi_{crit}$.
The evolution of the energy surface can be seen as the competition
between two minima, spherical and deformed, rather than a gradual evolution
from spherical
to weakly deformed to large deformation. Such a level crossing scenario is, in
fact,
characteristic of a first order phase transition.

The qualitative behavior of these results is not sensitive to
boson number N$_B$ nor to variations in $\chi$.  We show this in fig. 4b,
which
gives the values of the location of the lowest minimum of the energy surface,
 as a function of
$\xi$ for a set of N$_B$ and $\chi$ values.  The curves all show the
same behavior independent of the parameters:
$\beta_{min}$ is zero for small $\xi$ values, and then rises extremely rapidly
to a saturation value within a very narrow range of $\xi$ values (which
define  a $\xi_{crit}$ for each $N_B$ and $\chi$).

In fig. 4c, we show energy surfaces as a function of $\beta$ for IBM
parameters applicable to each of the Sm isotopes.  These surfaces range
from near vibrator shapes for $^{146,148}$Sm to softer in
$^{150}$Sm, to the unique coexistence nucleus $^{152}$Sm where
two shallow minima occur, to
$^{154,156}$Sm where the prolate minimum is much deeper,
thus localizing these latter nuclei at large positive $\beta$
values.  The actual minimum in the energy surface occurs {\it only}
for $\beta_{min}$ $\sim$ 0 or large positive $\beta_{min}$.

In summary, we have shown that finite nuclei can, and do, exhibit true phase
transitional character by proposing a way to resolve the integer
nucleon number problem.  We have shown that one can identify an empirical
quantity, E(2$^+_1$), that is nearly continuous and in terms of which other
quantities, such as E(4$^+_1$) or S$_{2n}$, follow simple, compact
trajectories with small fluctuations even for large regions of nuclei.  We
have
seen that these observables have distinct anomalies at similar
values of E(2$^+_1$) at which phase coexistence in
$^{152}$Sm has been recently identified.  Other observables such as
S$_{2n}$ may also be used to correlate structural changes.  Through a model,
such as the IBM
(the GCM gives similar results), we have associated the order parameter with a
physical
quantity, the deformation
$\beta_{min}$, at which the potential has a minimum.  In a spherical-deformed
transition
region in the IBM, $\beta_{min}$ takes on two characteristic values,
zero and then an abrupt change to a near saturation deformation.
It appears that nuclear structural evolution in this mass region entails two
basic phases (spherical and deformed) rather than a gradual
softening  (with valence nucleon number) traditionally associated with the
onset
of deformation in nuclei.  Although this view is
unconventional, the analysis of the IBM suggests that it is nearly
unavoidable and a basic feature of structural evolution.\\

We would like to thank F. Iachello, S. Kuyucak, V. J. Emery, S. Shapiro,
 J. Axe, and M.
Straayer for useful discussions of the concept of phase transitions in both
finite and infinite systems.

 Work supported under DOE Contract Numbers
DE-FG02-91ER-40609, DE-FG02-91ER-40608 and DE-FG02-88ER-40417.

\newpage

\newpage
{\bf Figure Captions}

\begin{enumerate}

\item[1:]  E(4$^+_1$) against neutron number for collective nuclei,
that is, nuclei for which
R$_{4/2}$ $\equiv$ E(4$^+_1$)/E(2$^+_1$) $>$ 2.05.
a) Sm, b) the 50
$<$ Z $\leq$ 66 region. Data from ref. [2]

\item[2:]  $E(4^+_1$) plotted against $E(2^+_1$).
Panels a-c), show the same data as
in fig. 1 for sequentially more elements.  Panel d) is an expanded view of the
rotational region, with all the data points from panel c) with
$E(2^+_1)<E_c(2^+_1)$.

\item[3:] (a) Separation energies $S_{2n}$ as a function of
  $E(2_1^+)$. S$_{2n}$ data from ref. [13].  (b) $B(E2:2_1^+\rightarrow
0_1^+)$
   values as a function
  of $S_{2n}$.  B(E2) data from ref. [14].

\item[4:] Classical limit analysis of IBM
calculations that reproduce fig. 2c.  a) Energy surface as a
function of $\xi$ [see eq. (2)] near the critical point for
N$_B$ = 10 and $\chi$ = -$\sqrt{7}$/2; b) location of the
minima, $\beta_{min}$, as a function of $\xi$, for several
values of N$_B$ and $\chi$; and c) The IBM energy E($\beta$) for
the $\xi$ values and boson numbers corresponding to the Sm
isotopes (see ref. [22]).  The minima occur only for $\beta$ = 0
or large finite values.  There is no gradual evolution of
$\beta$ from 0 to saturation levels. 

\end{enumerate}
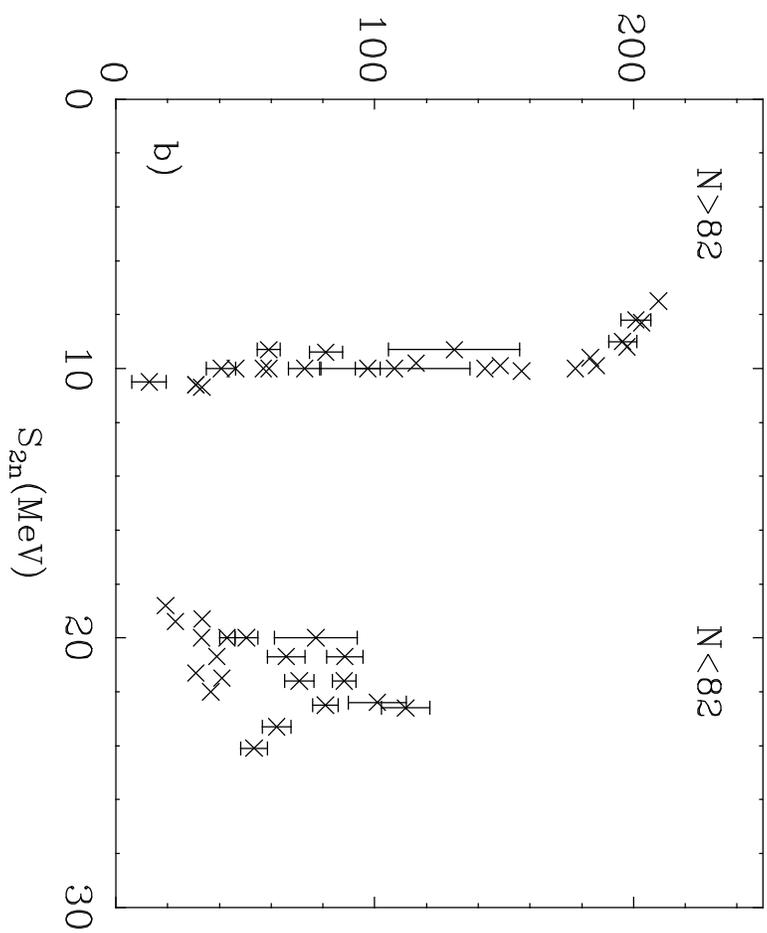
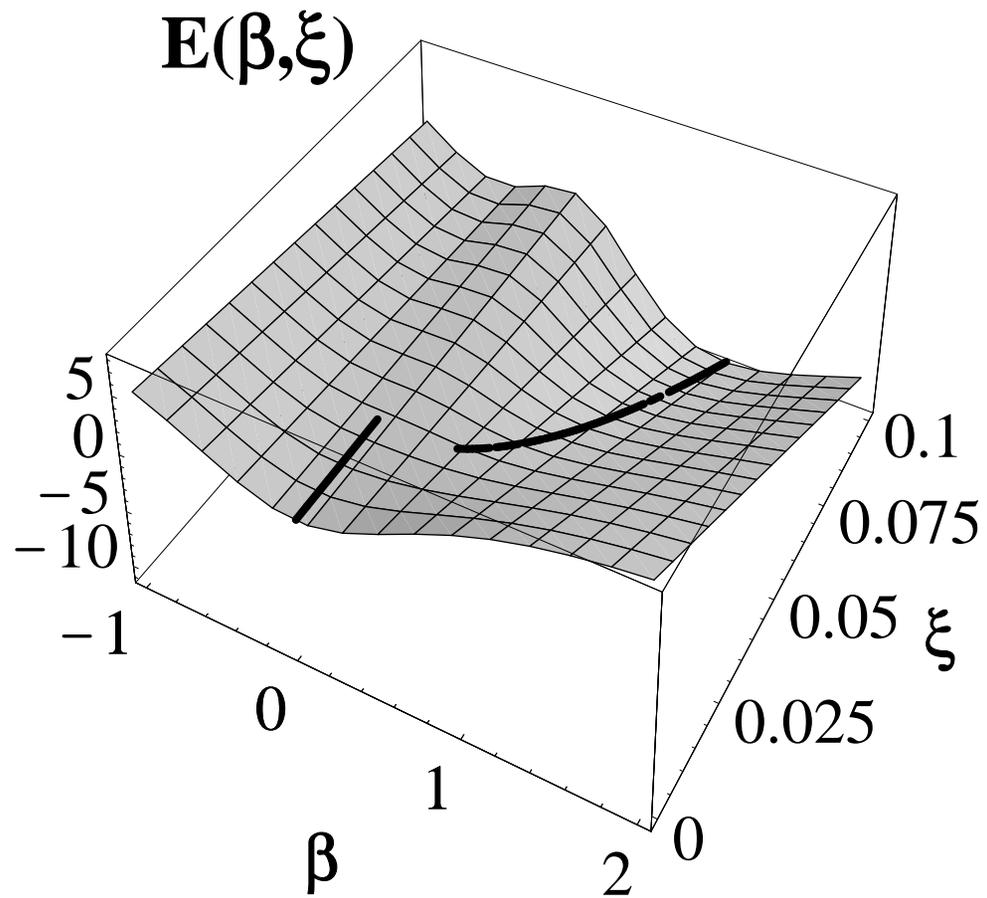
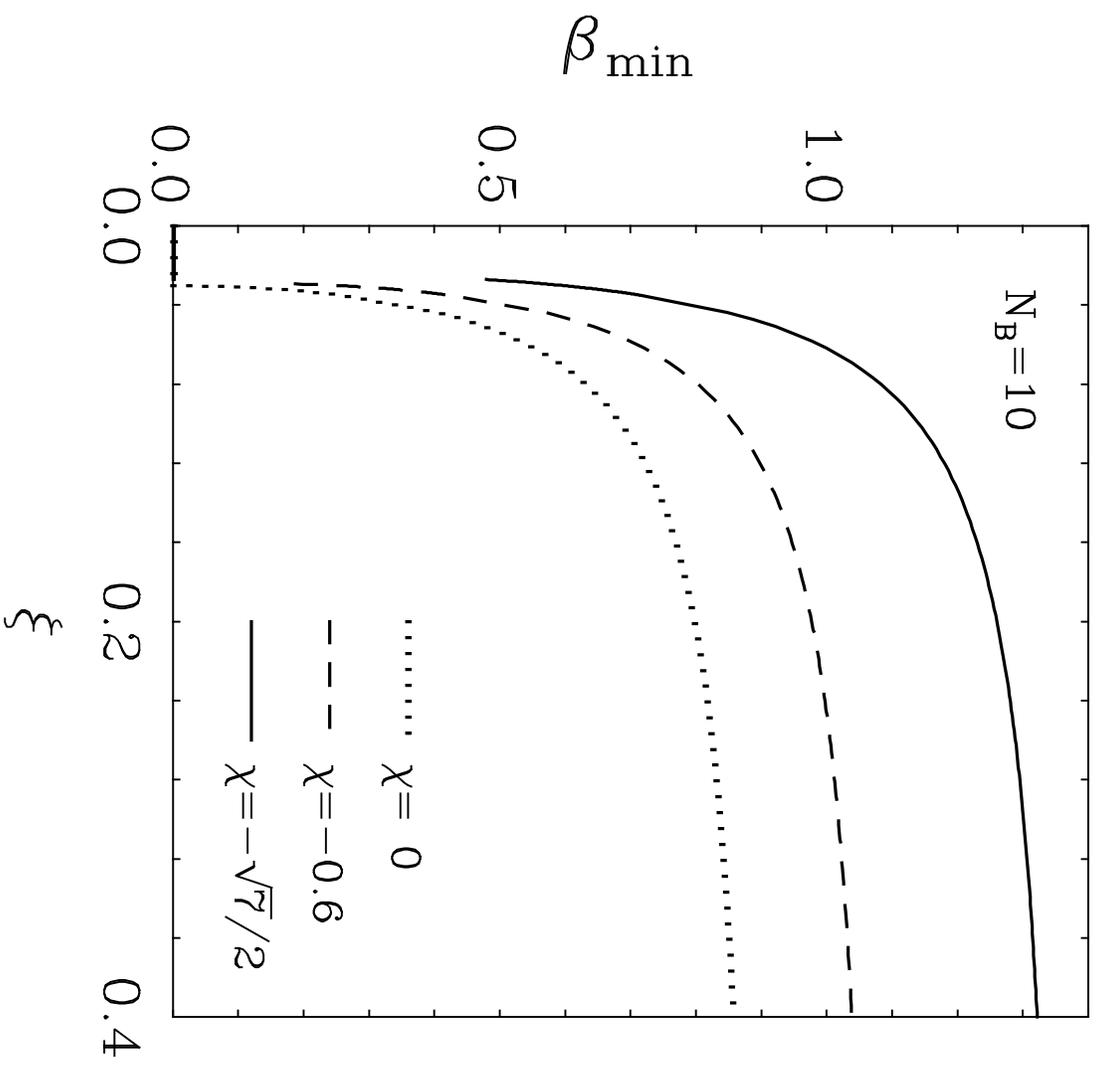
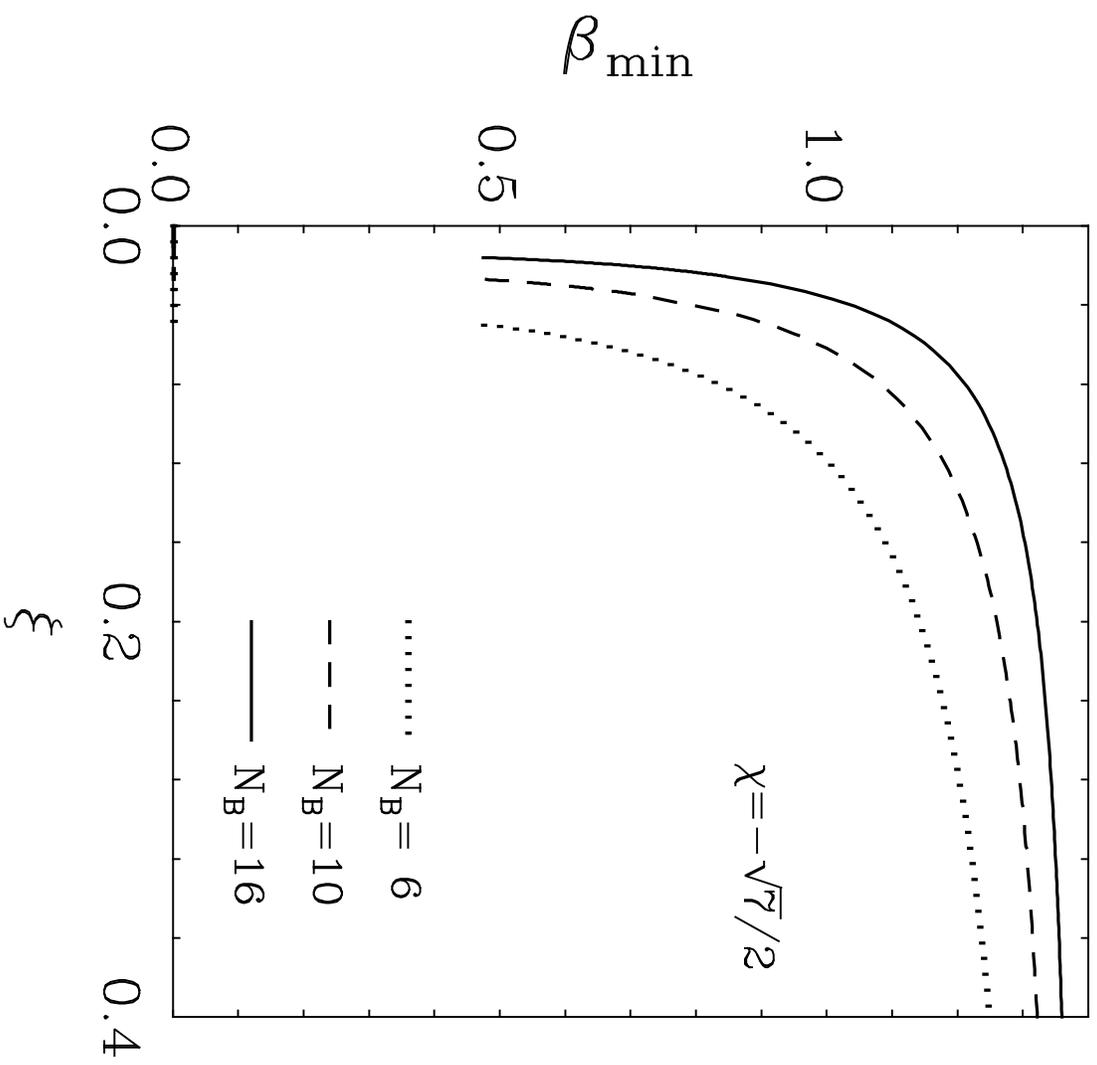
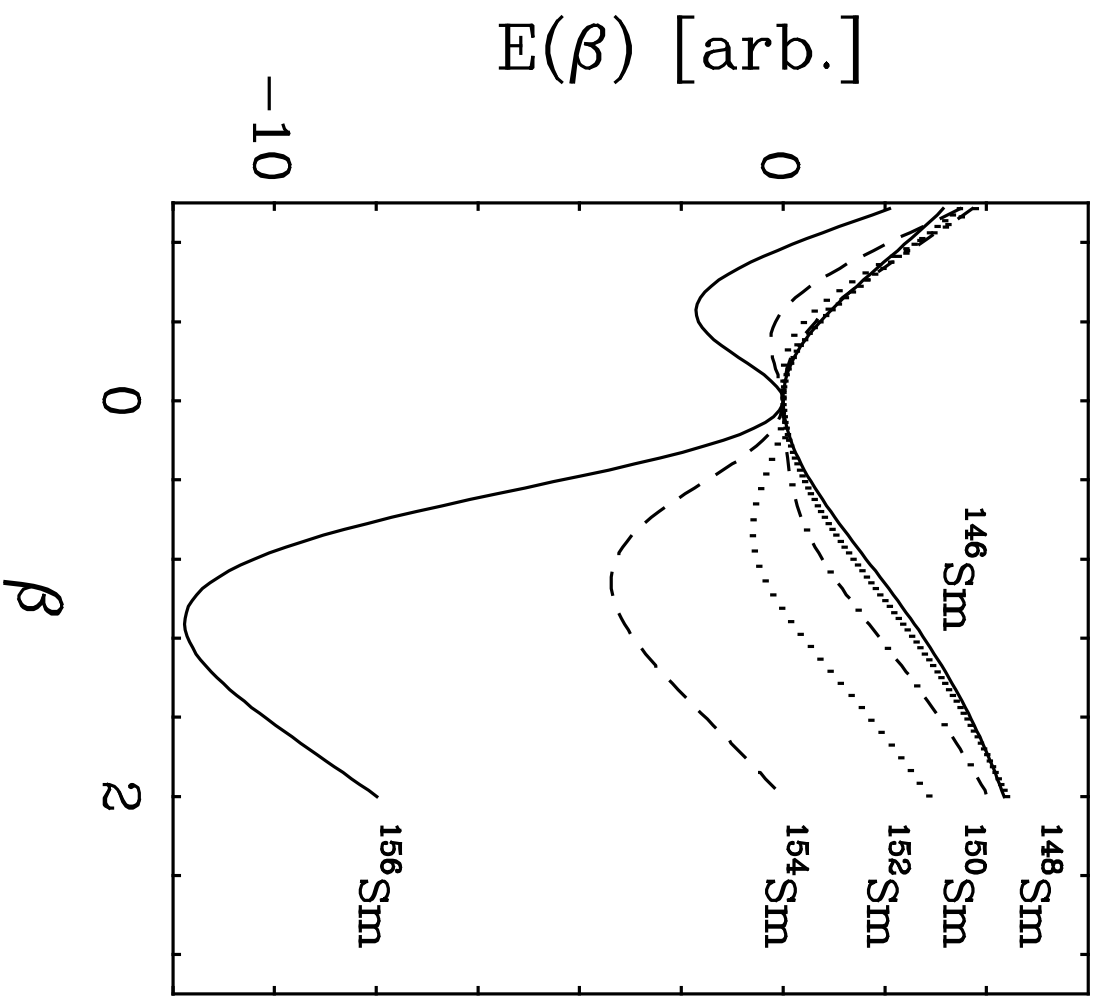
\end{document}